\documentclass[aps,pra,twocolumn,superscriptaddress,10pt]{revtex4-2}
\usepackage[utf8]{inputenc}
\usepackage{multirow,amsthm,amssymb,amsbsy,amsmath,epsfig,epstopdf,bm,float}
\usepackage{bm}
\usepackage{graphicx}
\usepackage{booktabs}
\usepackage{verbatim}
\usepackage{dcolumn}
\usepackage{color}
\usepackage{xcolor}
\usepackage[normalem]{ulem}
\usepackage{soul}
\usepackage{braket}
\usepackage{subfig}
\usepackage[export]{adjustbox}
\usepackage{enumerate}
\usepackage{hyperref}
\hypersetup{
    colorlinks=true,
    linkcolor=blue,
    citecolor=blue,
    urlcolor=blue
}
\usepackage{orcidlink}

\begin{document} 

\title{Measurement-enabled online quantum processing with amplitude encoding}
\author{{Giacomo Franceschetto}
\orcidlink{0009-0009-2200-9552}}
\email{giacomo.franceschetto@icfo.eu}
\affiliation{ICFO-Institut de Ciències Fotòniques, The Barcelona Institute of Science and Technology, 08860 Castelldefels (Barcelona), Spain}

\author{{Pere Mujal}
\orcidlink{0000-0002-4077-0304}}
\email{mujal@lorentz.leidenuniv.nl}
\affiliation{ICFO-Institut de Ciències Fotòniques, The Barcelona Institute of Science and Technology, 08860 Castelldefels (Barcelona), Spain}
\affiliation{Instituut-Lorentz, Universiteit Leiden, P.O. Box 9506, 2300 RA Leiden, The Netherlands.}
\affiliation{$\langle \textit{aQa}^\textit{L} \rangle$ Applied Quantum Algorithms, Leiden University, The Netherlands.}

\author{{Rodrigo Mart\'inez-Pe\~na}
\orcidlink{0000-0002-1459-7464}}
\email{rodrigo.martinez@dipc.org}
\affiliation{Donostia International Physics Center, Paseo Manuel de Lardizabal 4, E-20018 San Sebastián, Spain}

\date{ \today }

\begin{abstract}
We introduce a quantum reservoir computing online protocol that realizes amplitude encoding on quantum hardware. Our scheme combines mid-circuit measurement and reset operations to implement the partial-trace dynamics underlying amplitude encoding, while an indirect measurement scheme provides access to reservoir observables without interrupting temporal processing. In contrast to other approaches, our method preserves online operation, avoids input buffering, and keeps the runtime linear in the number of time steps. We present the theoretical formulation of the protocol and a proof-of-principle implementation on quantum hardware, and we evaluate its performance on two standard benchmark tasks. Our results show that the reservoir dynamics can be monitored through both direct measurements of the input qubits and indirect measurements of the memory qubits, enabling observation of the full system while isolating the internal evolution of the reservoir. This work provides a practical route toward scalable hardware implementations of amplitude-encoded quantum reservoir computing and opens the door to systematic experimental studies of complex quantum reservoirs.
\end{abstract}


\maketitle

\section{Introduction}
Quantum reservoir computing  (QRC) aims to use the natural dynamics of quantum systems as a computational resource for temporal information processing \cite{Neuromorphic2021rev,Opportunities2021rev}. The first QRC proposal, introduced by Fujii and Nakajima \cite{fujii2017harnessing}, realized this idea with an interacting spin reservoir. In this scheme, inputs are injected at each time step by tracing out the state of the encoding qubits while preparing a new state that is parametrized by the input. This amplitude encoding scheme has since become a reference model for QRC \cite{nakajima2019boosting, chen2019learning, kutvonen2020optimizing, tran2020higher, martinez2020information, martinez2021dynamical, nokkala2021gaussian, nokkala2021high, mujal2021analytical, xia2022reservoir, mujal2022quantum, vintskevich2022computing, xia2023configured, llodra2023benchmarking, mujal2023time, gotting2023exploring, mlika2023user, garcia2023scalable, domingo2023optimal, kobayashi2023quantum, nokkala2024retrieving, garcia2024squeezing, palacios2024role, vcindrak2024enhancing, kobayashi2024coherence, kobayashi2024extending, llodra2025quantum, hahto2025smarter, ivaki2025quantum, vcindrak2025engineering, vcindrak2025krylov, schutte2025expressive, steinegger2025predicting, kora2025statistical, sasaki2025hamiltonian, kobayashi2025quantum, kobayashi2025edge, sannia2025exponential, xiong2025role, askari2025spin, jain2026higher, kawanabe2026efficient, vcindrak2026memory, correlatedSun2026}.

Its popularity stems from the fact that nonlinear input codification \cite{mujal2021analytical} and state contraction (necessary for providing memory; see, for example,  \cite{martinez2021dynamical}) are combined into a single map. This makes it possible to explore complex quantum models for processing input information, isolating the dynamics of the reservoir for a better analysis of its role \cite{martinez2021dynamical,xia2022reservoir,palacios2024role,llodra2025quantum,kobayashi2025edge}. It has also enabled the study of universal approximation theorems \cite{chen2019learning,nokkala2021gaussian}, the exponential concentration phenomenon \cite{sannia2025exponential,xiong2025role}, and online protocols \cite{mujal2023time,franceschetto2026harnessing}. However, despite its popularity, real hardware implementations of this scheme remain elusive. The main difficulty is that each input update requires tracing out the state of the encoding qubits without interrupting the algorithm, while also monitoring the largest possible number of system elements.

\begin{figure*}[t]
\centering
\includegraphics[width=0.85\textwidth]{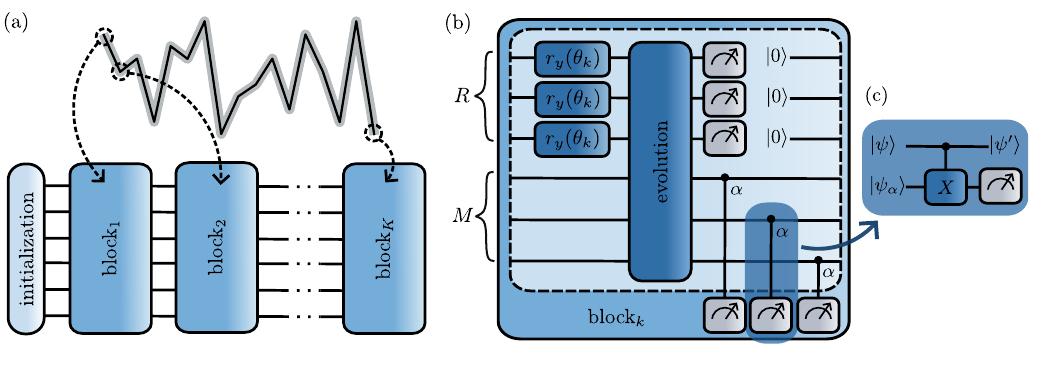}
\caption{Quantum reservoir computing scheme for the online protocol using amplitude encoding. (a) The complete circuit for the algorithm over $K$ time steps, with one block per step. (b) Description of a circuit block at time step $k$. (c) Description of the indirect measurement circuit for a memory qubit and its ancilla.}
\label{Fig:1}
\end{figure*}

In this work, we address this gap in the literature by introducing and testing an experimental online processing design that implements amplitude encoding. We provide the theory and a proof-of-principle implementation on quantum hardware, evaluating two standard benchmark tasks. Our design is based on two ideas, summarized in the scheme of Fig.\ref{Fig:1}.

First, we implement the amplitude encoding map through mid-circuit measurements and reset operations.  This idea is inspired by prior work on mid-circuit measurements in QRC \cite{hu2024overcoming}. There, the available qubits are split into memory and readout sets. Inputs are encoded with a unitary acting on all qubits, after which mid-circuit measurements extract information from the readout qubits whereas memory qubits preserve input information for later time steps. For our design, the crucial feature is that reset operations after measurement, averaged over outcomes, introduce the partial trace into the reservoir equations. This has recently been exploited to reproduce the amplitude encoding scheme on quantum hardware \cite{hamhoum2025multivariate}. However, this approach relies on rewinding protocols \cite{chen2020temporal,mujal2023time}, where the dynamics is reconstructed from sliding time windows rather than from a single online update rule; this makes the performance dependent on the window size and initial state \cite{vcindrak2024enhancing}, and requires the input sequence to be buffered \cite{mujal2023time}.

Second, we adopt an indirect-measurement protocol to access reservoir observables while preserving online processing, building on previous work \cite{mujal2023time,franceschetto2026harnessing,yasuda2023quantum, ricci2026quantum}. Our goal is to exploit observables from all qubits while keeping the experimental runtime linear in the number of time steps. When several copies of the system are available, as for atomic or molecular ensembles \cite{negoro2021toward}, expectation values of these observables can be computed in real time. To this end, we use ancillary qubits to indirectly extract observables from the memory qubits while controlling the associated measurement backaction. 

Our approach monitors the memory qubits through indirect measurements alongside projective measurements of the input qubits. This makes observables accessible across the entire system while isolating the internal dynamics of the reservoir into a separate channel. By making amplitude encoding and tunable measurement backaction experimentally accessible within the same online design, the protocol opens a route to study and exploit monitored quantum dynamics for temporal information processing.

\section{Model}We describe the evolution of a quantum reservoir in discrete time steps with a recurrent equation,
\begin{equation}
\label{eq:rhokunperturbed}
    \rho^{\textnormal{u}}_k=\mathcal{L}_k[\rho^{\textnormal{u}}_{k-1}],
\end{equation}
where $k$ labels the input sequence, the superscript u denotes unperturbed dynamics and $\mathcal{L}_k$ represents a completely positive and trace preserving (CPTP) quantum channel. This general formulation does not impose any restriction on the input codification.

Ideally, information would be extracted through the expectation value of any observable $\hat{\mathcal{O}}$ at each time step over an infinite ensemble, 
\begin{equation}
\label{eq:ideal_means}
    \langle \hat{\mathcal{O}} \rangle^{\infty}_{\rho_k^{\rm u}} :=\textnormal{Tr}\left(\hat{\mathcal{O}}\rho^{\textnormal{u}}_k \right).
\end{equation}
In experiments, this expectation value is estimated from $N_{\rm meas}$ measurements on identical copies, and is therefore subject to statistical error.

Let us now consider the case of introducing measurements in the reservoir equations. For a single run of the online protocol, the monitored state  $\rho_{\mathbf{i}_k}$ (for which we drop the apex `u') evolves following a quantum trajectory characterized by the  (stochastic) measurement outcome $\mathbf{i}_k$:
\begin{equation}
\label{eq:rhotrajectories}
    \rho_{\mathbf{i}_k}=\frac{\left(\mathcal{M}_{\mathbf{i}_k} \circ  \mathcal{L}_k\right)[\rho_{\mathbf{i}_{k-1}}]}{\textnormal{Tr}\left(\left(\mathcal{M}_{\mathbf{i}_k} \circ  \mathcal{L}_k\right)[\rho_{\mathbf{i}_{k-1}} ]\right)},
\end{equation}
where the effect of the measurements on the system is determined by $\mathcal{M}_{\mathbf{i}_k}$. We use measurement effects of the form $\hat{E}_{\mathbf{i}_k} = \hat{\Omega}^\dagger_{\mathbf{i}_k}\hat{\Omega}_{\mathbf{i}_k}$, with Kraus operators $\hat{\Omega}_{\mathbf{i}_k}$ specified below. The outcome $\mathbf{i}_k$ is a random vector in $\{0,1\}^R$, where $R$ is the number of directly measured qubits, which also serve as input-reset qubits. The remaining $M=:N-R$ qubits will be denoted as memory qubits.

Assuming an infinite ensemble of realizations, the mixed state that accounts for all possible measurement outcomes at time step $k$ is given by:
\begin{equation}\label{eq:rhok}
\rho_k=\sum_{\mathbf{i}_k}\rho_{\mathbf{i}_{k}}P(\mathbf{i}_k).
\end{equation}

\subsection{Amplitude encoding with measurement and reset}
Let us consider the following Kraus measurement operator:
\begin{equation}\label{eq:meas_op}
   \hat{\Omega}_{\mathbf{i}_k} = \ket{\mathbf{0}}\bra{\mathbf{i}_k}\otimes \hat{I}_{M},
\end{equation}
where $\ket{\mathbf{0}}$ is the zero state for $R$ qubits, $\ket{\mathbf{i}_k}$ is the state of the $R$-measured qubits, and $\hat{I}_M$ is the identity operator for the $M$ memory qubits. 
This operator is a combination of the projection and reset operations of the $R$  qubits under measurement, and it can be constructed by the concatenation of the projection and reset operator of each individual measured qubit (in any order, since they commute). Applying the definition of marginal probability and conditional probability over $P(\mathbf{i}_k)$, Eq.~\eqref{eq:rhok} becomes:
\begin{equation}
 \begin{split}
     \rho_k&=\sum_{\mathbf{i}_k}\rho_{\mathbf{i}_k}P(\mathbf{i}_k) = \sum_{\{\mathbf{i}_k,\dots,\mathbf{i}_1\}}\rho_{\mathbf{i}_k}P(\mathbf{i}_k,\dots,\mathbf{i}_1) \\
     &= \sum_{\{\mathbf{i}_k,\dots,\mathbf{i}_1\}}\rho_{\mathbf{i}_k}P(\mathbf{i}_k|\mathbf{i}_{k-1},\dots,\mathbf{i}_1)P(\mathbf{i}_{k-1},\dots,\mathbf{i}_1).
 \end{split}
\end{equation}
Now, note that the denominator in Eq.~\eqref{eq:rhotrajectories} is the probability $P(\mathbf{i}_k|\mathbf{i}_{k-1},\dots,\mathbf{i}_1)$, since it quantifies the probability of obtaining the outcome $\mathbf{i}_k$ given the previous record of outcomes, encoded in the conditional state $\rho_{\mathbf{i}_{k-1}}$. Substituting:
\begin{equation}
 \begin{split}
     \rho_k&= \sum_{\{\mathbf{i}_k,\dots,\mathbf{i}_1\}}\left(\mathcal{M}_{\mathbf{i}_k} \circ  \mathcal{L}_k\right)[\rho_{\mathbf{i}_{k-1}}]P(\mathbf{i}_{k-1},\dots,\mathbf{i}_1)\\
     & = \sum_{\mathbf{i}_k} \hat{\Omega}_{\mathbf{i}_k}\mathcal{L}_k\left[\sum_{\{\mathbf{i}_k,\dots,\mathbf{i}_1\}}\rho_{\mathbf{i}_{k-1}}P(\mathbf{i}_{k-1},\dots,\mathbf{i}_1)\right]\hat{\Omega}^\dagger_{\mathbf{i}_k}\\
     & = \sum_{\mathbf{i}_k} \hat{\Omega}_{\mathbf{i}_k}\mathcal{L}_k\left[\rho_{k-1}\right]\hat{\Omega}^\dagger_{\mathbf{i}_k},
 \end{split}
\end{equation}
where we have used the linearity of $\mathcal{M}_{\mathbf{i}_k} \circ  \mathcal{L}_k$ in the first equality. Let us define $\tilde{\rho}_{k}= \mathcal{L}_k\left[\rho_{k-1}\right]$. Applying Eq.~\eqref{eq:meas_op} to the previous equation, we obtain $\rho_k = \ket{\mathbf{0}}\bra{\mathbf{0}}\otimes \text{Tr}_R\left(\tilde{\rho}_{k}\right)$,
and if we apply the map $\mathcal{L}_{k+1}$, we find
\begin{equation}
    \tilde{\rho}_{k+1}=\mathcal{L}_{k+1}[\rho_k] =\mathcal{L}_{k+1}\left[ \ket{\mathbf{0}}\bra{\mathbf{0}}\otimes \text{Tr}_R\left(\tilde{\rho}_{k}\right)\right].
\end{equation}
To recover the amplitude encoding scheme, we use an encoding map that only affects the measured qubits. For that, we decompose the CPTP map $\mathcal{L}_{k+1}$ in two parts, $\mathcal{L}_{k+1} := \mathcal{E}^{(R,M)}\circ \mathcal{J}^{(R)}_{k+1}$. The CPTP map $\mathcal{J}_{k+1}^{(R)}$ is the encoder over the measured qubits, and $\mathcal{E}^{(R,M)}$ is the entangler. Then, the reservoir map can be written as
\begin{equation}
\begin{split}
    \tilde{\rho}_{k+1}&=\mathcal{E}^{(R,M)}\left[ \mathcal{J}^{(R)}_{k+1}[\ket{\mathbf{0}}\bra{\mathbf{0}}]\otimes \text{Tr}_R\left(\tilde{\rho}_{k}\right)\right]\\
    &=\mathcal{E}^{(R,M)}\left[ \rho^{(R)}_{k+1}\otimes \text{Tr}_R\left(\tilde{\rho}_{k}\right)\right],
\end{split}
\end{equation}
which is the reservoir equation proposed by Fujii and Nakajima in \cite{fujii2017harnessing} when $R=1$, $\mathcal{J}_{k+1}^{(R)} = r_y(\theta_{k+1})$ is a rotation map in the $y$ axis, and  $\theta_{k+1}=2\arccos(\sqrt{1-s_{k+1}})$ is the rotation angle \cite{sannia2024dissipation}. 

Finally, we can calculate the expected value of the measured observables by evaluating them over an infinite number of trajectories; we use the first qubit as an example. From the definition of the marginal probability, we obtain the following expression for the observable $\hat{\mathcal{O}}_1 := \sum_{i^{(1)}}i^{(1)} \ket{i^{(1)}}\bra{i^{(1)}}\otimes \hat{I}_{R+M-1} $:
\begin{equation}\label{eq:O}
\begin{split}
        \braket{\hat{\mathcal{O}}_1}_k = \sum_{i^{(1)}_k}i^{(1)}_k P(i^{(1)}_k) = \sum_{\{\mathbf{i}_k,\dots,\mathbf{i}_1\}}i^{(1)}_k P(\mathbf{i}_k,\dots,\mathbf{i}_1).
\end{split}
\end{equation}
The probability of obtaining the list of outcomes $(\mathbf{i}_k,\dots,\mathbf{i}_1)$ in a trajectory starting from some initial condition $\rho_0$ is computed as $P(\mathbf{i}_k,\dots,\mathbf{i}_1)=\text{Tr}\left((\mathcal{M}_{\mathbf{i}_k}\circ \mathcal{L}_k)\circ \cdots \circ (\mathcal{M}_{\mathbf{i}_1}\circ \mathcal{L}_1) [\rho_0]\right)$. Using the linearity of CPTP maps and the completeness relation $\sum_{\mathbf{i}}\ket{\mathbf{i}}\bra{\mathbf{i}} = \hat{I}$, we can substitute the previous formula into Eq.~\eqref{eq:O} to arrive at the final expression of the observable:
\begin{equation}
\braket{\hat{\mathcal{O}}_1}_k=\text{Tr}\left(\hat{\mathcal{O}}_1\tilde{\rho}_{k}\right).\end{equation}
\subsection{Introducing indirect measurements}We introduce indirect measurements of the memory qubits during online processing \cite{mujal2023time,franceschetto2026harnessing} using $A$ ancillae as probes. Each measured memory qubit is entangled with one ancilla, giving $N+A$ qubits, with $A=M$ if all memory qubits are monitored. The probe qubits are then measured and reset, as for the input qubits. 

The Kraus operator in Eq.\eqref{eq:meas_op} is replaced by
\begin{equation}\label{eq:meas_op_weak}
   \hat{\Omega}_{\mathbf{n}_k} = \ket{\mathbf{0}}_R\bra{\mathbf{i}_k}\otimes \hat{I}_{M}\otimes \ket{\mathbf{0}}_A\bra{\mathbf{j}_k},
\end{equation}
where we defined the concatenated output vector $\mathbf{n}_k:=(\mathbf{i}_k,\mathbf{j}_k)$. Now we can rewrite the equations of the previous section for the new output vector $\mathbf{n}_k$, since adding the ancillae is equivalent to increasing $R$:
\begin{equation}
    \rho_k = \ket{\mathbf{0}}_R\bra{\mathbf{0}}\otimes \text{Tr}_{R,A}\left(\tilde{\rho}_{k}\right)\otimes  \ket{\mathbf{0}}_A\bra{\mathbf{0}}.
\end{equation}
However, we must define a new quantum channel that entangles each memory qubit with an ancilla: 
$\mathcal{L}_{k+1} :=\mathcal{A}^{(M,A)} \circ 
\mathcal{E}^{(R,M)}\circ \mathcal{J}^{(R)}_{k+1}$, with reservoir state
\begin{equation}
    \tilde{\rho}_{k+1}=\mathcal{A}^{(M,A)}\left[\mathcal{E}^{(R,M)}\left[ \rho^{(R)}_{k+1}\otimes \text{Tr}_{R,A}\left(\tilde{\rho}_{k}\right)\right]\otimes  \ket{\mathbf{0}}_A\bra{\mathbf{0}}\right].
\end{equation}
The indirect measurement strength is controlled by a CNOT gate and a rotation gate acting on every pair of memory qubit $m$ and ancilla qubit $a$, and the output information is obtained by measuring the ancillae, which are reset subsequently. Related ancilla-based measurement schemes have been proposed in \cite{chen2020temporal,yasuda2023quantum,ricci2026quantum,polche2026hybrid}.
Specifically, the interaction map implemented in every pair $(m,a)$ is the following:
\begin{equation}\label{eq:twoqubitindirect}
    \mathcal{A}^{(m,a)}(\rho_m\otimes\ket{0}_a\bra{0})= \text{CNOT}(\rho_m,r_y(\beta)\ket{0}_a\bra{0}),
\end{equation}  
where the $\text{CNOT}(c,t)$ gate acts on the memory qubit as control and the ancilla qubit as target, and the angle $\beta:=2\arccos(\alpha)$ is defined such that $r_y(\beta)\ket{0}_a\bra{0} = \ket{\psi_\alpha}\bra{\psi_\alpha}$, with $\ket{\psi_\alpha}=\alpha\ket{0}+\sqrt{1-\alpha^2}\ket{1}$ and $\alpha\in(1/\sqrt{2},1]$. The measurement strength is quantified by the value of $\alpha$, reaching the projective case for $\alpha=1$ and approaching the weak measurement limit when $\alpha\rightarrow 1/\sqrt{2}$.

This allows us to obtain the expected value $\langle \hat{\sigma}^z\rangle_m$ of the memory qubit through the one of the ancilla, $\langle \hat{\sigma}^z\rangle_a$, as $\langle \hat{\sigma}^z\rangle_m=\langle \hat{\sigma}^z\rangle_a/(2\alpha^2-1).$
Rotations before measurements are required if different axes are measured \cite{mujal2023time}.
The resulting Kraus operators describing the indirect measurement process acting on the memory qubit are
\begin{equation}
    \hat{\Omega}_s=\sqrt{\delta_{s,-1}+s\alpha^2}\ket{0}\bra{0}+\sqrt{\delta_{s,1}-s\alpha^2}\ket{1}\bra{1},
\end{equation}
with $s \in\{-1,1\}$ and $\delta_{s,s'}$ the Kronecker delta. The unconditional state of the memory qubit after the measurement is
\begin{equation}\label{eqMeasancilla}
    \rho'_m=\hat{M}_m\odot\rho_m, \ \hat{M}:= \hat{I}+2\alpha\sqrt{1-\alpha^2}\hat{\sigma}^x,
\end{equation}
where $\odot$ is the Hadamard or element wise matrix product. Equation \eqref{eqMeasancilla} shows the equivalence between the present dephasing channel, induced by discrete-variable ancilla measurements in qubit circuits, and the continuous-variable ancilla measurements previously defined in spin systems \cite{mujal2023time,franceschetto2026harnessing}. In the latter case, the dephasing term corresponds to $ 2\alpha\sqrt{1-\alpha^2}=e^{-g^2/2}$, where $g$ is the measurement strength. Sharp projective measurements are given by $g\gg 1$, while weak measurements correspond to $g\ll 1$. In the following, this relation between $g$ and $\alpha$ is always assumed, and we will use $g$ to refer to the measurement strength.

\begin{figure}[t!]
\captionsetup[subfigure]{}
\begin{center}
\includegraphics[scale = 0.9]{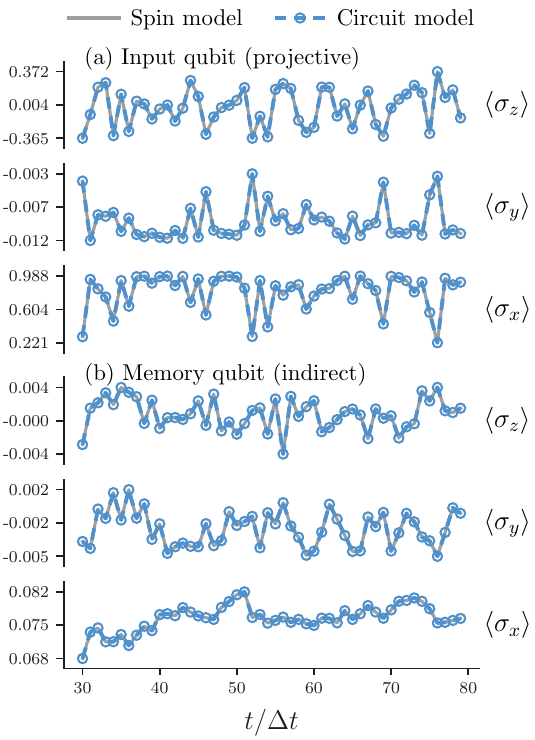}
\caption{
Comparison of expectation values obtained from the spin model and the circuit model for (a) the input qubit under projective measurement and for (b) a memory qubit under indirect measurement. 
}\label{Fig:2}
\end{center}
\end{figure}
\section{Implementation and validation}
\subsection{Reproducing the spin model}The original proposal of the online protocol focused on quantum analog simulation platforms with access to indirect measurements \cite{mujal2023time,franceschetto2026harnessing}. As a first step and for the sake of completeness, we numerically compare the dynamics of the original proposal and the quantum circuit model introduced here. 

We consider a quantum reservoir consisting of a spin network that evolves over time under the influence of the disordered transverse-field Ising Hamiltonian \cite{fujii2017harnessing}:
\begin{equation}
\hat{H}=\frac{1}{2}\sum_{i=1}^{N}h\hat{\sigma}_i^{z}+\sum_{i=1<j}^{N}J_{ij}\hat{\sigma}_i^{ x}\hat{\sigma}_j^{x},
\label{hamtransverseising}
\end{equation}
 where couplings $J_{ij}$ are drawn from a random uniform distribution in the interval $[-J_s/2,J_s/2]$, $h$ is an homogeneous external magnetic field, and $N$ is the number of qubits. The evolution after input injection is given by the CPTP map \begin{equation}
     \tilde{\rho}_{k+1}=e^{-i\hat{H}\Delta t} \left[ \rho^{(R)}_{k+1}\otimes \text{Tr}_R\left(\tilde{\rho}_{k}\right)\right]e^{i\hat{H}\Delta t},
 \end{equation}
 where $\Delta t$ determines the duration of the unitary dynamics. For the spin model proposed in \cite{mujal2023time}, averaging over an infinite ensemble of indirect measurements on the $z$ axis results in the following state:
 \begin{equation}
          \rho_k =\hat{M}^{\otimes N}\odot \tilde{\rho}_{k}=\hat{M}^{\otimes N}\odot \mathcal{L}_k[\rho_{k-1}].
 \end{equation}
In order to compare with the circuit model proposed above, we will replace $\hat{M}^{\otimes N}$ with $\hat{I}\otimes\hat{M}^{\otimes (N-1)}$ such that the first qubit is projectively measured and the remaining qubits are measured indirectly with the same $g$ value.

The equivalence between the models is illustrated in Fig.~\ref{Fig:2}, displaying a numeric comparison of the spin network (gray continuous line) and circuit simulation (light blue dashed line with circles). It shows the expectation values for an input qubit under a projective measurement (top panel) and a memory qubit under an indirect measurement (bottom panel).  Inputs are randomly drawn from a uniform distribution in $[0,1]$. The circuit model results are obtained using the Qiskit density-matrix simulation backend, and therefore are free from shot noise and no Trotterization of the unitary dynamics is required, enabling direct benchmarking. Data are shown for a reservoir composed of one input qubit and five memory qubits ($R=1$, $M=5$) with parameters $h=0.1$, $J_s = 1$, $\Delta t = 10$ and $g = 0.256$. The value of $g$ is chosen as a reference value from \cite{franceschetto2026harnessing}, where it was found to be optimal for solving the linear memory task, as also shown below in Fig.~\ref{Fig:g_app}.
Expectation values are shown for the three measurement directions, $\langle \hat{\sigma}^{x} \rangle$, $\langle \hat{\sigma}^{y} \rangle$, and $\langle \hat{\sigma}^{z} \rangle$, demonstrating perfect agreement between the two descriptions of the dynamics. 

\subsection{Hardware-tailored experiment}To validate our proposal, we performed a proof-of-principle experiment emulating the benchmarking of \cite{franceschetto2026harnessing}. Our proposal is not limited in terms of system scalability and input series length. However, some adaptations were required due to hardware limitations encountered during implementation.  We employed the \texttt{ibm\_basquecountry} machine, which is a Heron r2 IBM system with 156 qubits. 

\begin{figure}[t!]
\captionsetup[subfigure]{}
\begin{center}
\includegraphics[scale = 1.1]{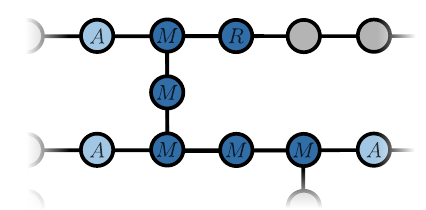}
\caption{Nine-qubit layout used for the hardware implementation, comprising one input/readout qubit (\(R\)), five memory qubits (\(M\)), and three ancilla qubits (\(A\)).
}\label{Fig:hardware_FN}
\end{center}
\end{figure}

Due to the heavy-hexagonal lattice topology of the quantum processor, simulating an all-to-all connected Hamiltonian would significantly increase gate errors due to the SWAP operations required for communication. We therefore switch from all-to-all connections to the nearest-neighbor transverse-field Ising model, eliminating the need of SWAP gates for the Hamiltonian simulation. The layout (Fig. \ref{Fig:hardware_FN}) also keeps the monitored memory qubits and their ancillae as nearest neighbours.  We use a 9-qubit layout (including ancillae), in which the first qubit is used for input and readout ($R = 1$) and five qubits are used for memory ($M = 5$). However, only three of these memory qubits are monitored via an ancilla due to the neighboring restriction. We specify this 9-qubit connectivity pattern as an initial layout to avoid SWAP insertion, and then let the Qiskit transpiler optimize its placement on the device, selecting the physical region that minimizes gate and readout errors.

Another limiting factor of the hardware implementation is the length of the classical register, which restricts individual hardware circuits to approximately 400 time steps in our implementation. Therefore, for usual RC tasks involving thousands of inputs, we need to divide the algorithm into pieces. In our implementation, we do this using a rewinding  procedure, restricting each hardware run to windows of 368 data points. This windowing is imposed by the present hardware rather than by the proposed online protocol. To remove the dependence on the initial state of each window, consecutive windows overlap by 50 points, providing an effective washout phase while preserving the relevant temporal context across window boundaries. However, the chain topology of the Hamiltonian exhibits a wide range of convergence rates (not shown). In our work, random seeds are selected such that the corresponding reservoir dynamics have an effective washout phase of about 50 time steps.

We additionally use dynamical decoupling for idle-qubit error suppression during mid-circuit measurements and fractional-gate compilation to reduce circuit depth. Altogether, these changes allow us to perform our experiment with a manageable error. 

Our benchmark comprehends two well-known tasks: the forecasting of a chaotic time series, represented by the Santa Fe task \cite{PhysRevA.40.6354,298828}; and the storage of past information, represented by the short-term memory (STM) task \cite{jaeger2001short}. The Santa Fe target is defined as $y_k = s_{k+\eta}$, where $\eta$ is the prediction distance and whose input data can be found in \cite{git_tuto}. The STM target is defined as $y_k = s_{k-\eta}$, where $\eta$ now represents the required memory, and the inputs are randomly and uniformly distributed within the interval [0, 1].  We use input sequences of 1000 points for the STM task and 1952 points for the Santa Fe task. We split the dataset into a training set and a test set comprising $70\%$ and $30\%$ of the data, respectively.
 
Given $\eta$, the capacity of the reservoir to perform each task can be measured as \begin{equation}
    C(\eta) = \frac{\text{cov}^2(\textbf{y},\hat{\textbf{y}})}{\text{var}(\textbf{y})\text{var}(\hat{\textbf{y}})},
\end{equation}
where $\textbf{y}$ is the test target sequence and $\hat{\textbf{y}}$ is the test predicted sequence. 

\begin{figure}[b!]
\captionsetup[subfigure]{}
\begin{center}
\includegraphics[scale = 0.9]{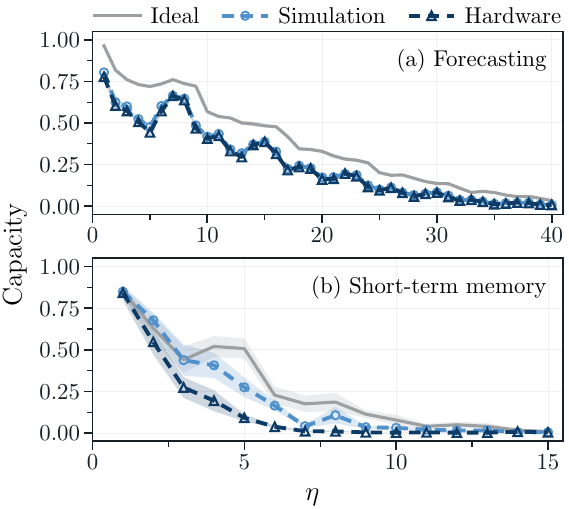}
\caption{Capacity of the hardware-tailored reservoir architecture for (a) predicting the Santa Fe time series and (b) solving the STM task.
}\label{Fig:3}
\end{center}
\end{figure}

Figure \ref{Fig:3} shows the capacities $C(\eta)$ for our experiment. Ideal (gray continuous line) denotes results from the ideal spin model simulation (no shot noise); Simulation (light blue dashed line with circles) corresponds to the circuit implementation with Qiskit software (\texttt{AerSampler()} backend), using first-order Trotter approximation with 10 steps, and $10^4$ shots for measurements. Hardware results (the dark blue dashed line with triangles) use the same Trotter approximation, steps and number of shots. Curves show averages over 10 random realizations of the Hamiltonian, and shaded regions indicate standard deviations of the means. Both tasks use a nearest-neighbor Hamiltonian with $h=1$, $J_s=1$, $\Delta t=10$, $R=1$, and $M=5$, while the output layer is made of the observables $\langle \hat{\sigma}^{x,y,z} \rangle$ of the readout qubit and three memory qubits (see Appendix \ref{app:B} for details). Similar to the optimization of measurement strength in \cite{franceschetto2026harnessing}, the value of $g$ (and thus $\alpha$) is optimized for each task. The values of $g$ are 2.05 and 0.256 for the forecasting and STM tasks, respectively. See Fig.~\ref{Fig:g_app} in Appendix \ref{app:A} for more details.

Figure \ref{Fig:3}(a) shows very good agreement between the circuit simulation and hardware result, both close to the ideal spin model line. The remaining gap is due to Trotterization and finite shot errors, while hardware noise plays a minor role. In Fig.~\ref{Fig:3}(b), differences appear around $\eta = 5$, although the overall trend is preserved. Here, hardware noise plays a more significant role, as discussed through Figs.~\ref{Fig:dyn_app} and \ref{Fig:C_combinations} in the Appendix \ref{app:D}. 

\section{Discussion and conclusion} We have shown that the amplitude-encoding scheme introduced by Fujii and Nakajima \cite{fujii2017harnessing} can be realized as an online dynamic-circuit protocol. The key step is to identify mid-circuit measurement followed by reset as the hardware primitive that implements the partial trace required by input injection. Combining this with ancillary indirect measurements gives access to reservoir observables while controlling the associated measurement backaction. In this sense, the protocol provides a hardware realization of two central ingredients of online QRC: amplitude encoding, which isolates the reservoir dynamics, and tunable monitoring, which can enhance temporal processing \cite{mujal2023time,franceschetto2026harnessing}.

The different impact of hardware noise on the two benchmark tasks reflects both the measurement regime and how information is distributed among the measured observables. Forecasting is performed close to the projective-measurement regime and is comparatively robust, whereas the STM task operates in the weak-measurement regime, where the $x$ and $y$ observables of the indirectly measured qubits are more strongly degraded by hardware noise. Although most of the STM capacity is carried by the more resilient $z$ observables, the $x$ and $y$ observables of the indirectly measured qubits also contribute to the task. This explains why hardware noise affects STM more strongly without preventing the reservoir from retaining substantial memory. A detailed analysis of this behavior is provided in Appendix \ref{app:D}.

To perform our design, we require a dynamical platform where mid-circuit measurements and reset operations are available. Our proof-of principle experiment is performed on an IBM quantum platform with superconducting qubits, where dynamic programming is possible \cite{corcoles2021exploiting}, while trapped-ion quantum computers are also available \cite{pino2021demonstration,decross2023qubit,yu2025situ}.  However, the entangling unitary dynamics of the model needs to be Trotterized. Further experimental platforms include dynamic programming and could be employed in the future to directly simulate the entangling dynamics without Trotterization, like arrays of neutral atoms \cite{graham2023midcircuit,lis2023midcircuit,norcia2023midcircuit}, or trapped ion simulators \cite{motlakunta2023preserving,foss2024progress}.

More broadly, our results show how measurement, reset, and controlled backaction can be engineered into the dynamics of a temporal quantum processor. This opens a route to experimentally study complex quantum reservoirs as monitored dynamical systems and to exploit their dynamics for online information processing.

\section*{Acknowledgements}We thank Luciano Pereira and Javier Oliva del Moral for valuable discussions and guidance regarding IBM Quantum hardware, and Antonio Acín, Marcin Płodzień, Oriol Morguí, Gian Luca Giorgi, and Roberta Zambrini for helpful discussions. The authors acknowledge the computational resources and technical support provided by the BasQ Strategy, under the collaboration agreement between Ikerbasque Foundation and Donostia International Physics Center (DIPC), on behalf of the Department of Science, Universities and Innovation of the Basque Government. G. F. acknowledges support from “la Caixa” Foundation (ID100010434) fellowship with code LCF/BQ/DI23/11990070.
P. M. acknowledges support from the European Union through a Marie Skłodowska-Curie Actions Postdoctoral Fellowship (Grant Agreement No. 101273006, DREAM-QA). Funded by the European Union. Views and opinions expressed are, however, those of the author(s) only and do not necessarily reflect those of the European Union or the European Research Executive Agency. Neither the European Union nor the granting authority can be held responsible for them.
R. M. P. acknowledges the QCDI project funded by the Spanish Government.
This project has received funding from  the Government of Spain (Severo Ochoa CEX2019-000910-S, FUNQIP and QEC4QEA PCI2025-163167), Fundació Cellex, Fundació Mir-Puig, Generalitat de Catalunya (CERCA program) and European Union (NextGenerationEU PRTR-C17.I1,
PASQuanS2.1, 101113690 and QEC4QEA, 101194322).

\section*{Data availability}The data and codes that support the findings of this article are
openly available \cite{git_tuto, franceschetto_2026_20399163}.
\section*{Author contributions}
R. Martínez-Peña and P. Mujal developed the research idea and derived the analytical results.  G. Franceschetto programmed and performed the numerical simulations and hardware experiment. All authors discussed the results and contributed to the writing of this manuscript.

\appendix
\section{Indirect measurement optimization}\label{app:A}
\begin{figure}[b]
\captionsetup[subfigure]{}
\begin{center}
\includegraphics[scale = 0.8]{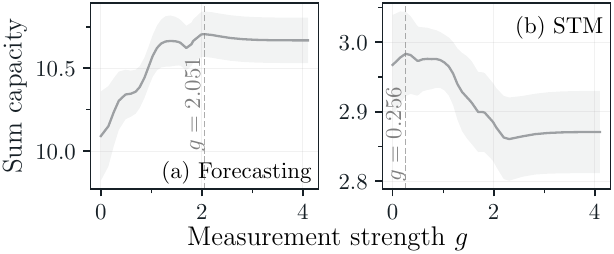}
\caption{
Optimization of the measurement strength with the spin model for the hardware-tailored experiment.
}\label{Fig:g_app}
\end{center}
\end{figure}

Similar to \cite{franceschetto2026harnessing}, we search the optimal values of the measurement strength for each task in Fig.~\ref{Fig:g_app}. To efficiently perform the optimization, we used the spin model as a proxy of the hardware-tailored architecture to identify the optimal operating point in the (a) forecasting task and (b) STM task. We use the mean summed capacity $\sum_{\eta} C(\eta)$ as our metric, averaged over 10 random realizations. We plot it as a function of the measurement strength $g$, with shaded regions indicating the standard error over realizations. Data are obtained using the same parameters as in the main text. Vertical dashed lines indicate the optimal measurement strengths selected for the hardware and circuit-model results discussed in Fig.~\ref{Fig:3}.

Looking at the vertical axis of Fig.~\ref{Fig:g_app}, we can see that the effect of $g$ on an output layer consisting of single-qubit observables is not very large. In fact, we found that hardware realisations with different $g$ values ($\alpha$) produced similar results. However, observables with a larger support, such as two-qubit correlations, can exhibit greater sensitivity, as demonstrated in Fig.~4(a) of Ref. \cite{mujal2023time}, highlighting the importance of optimizing the measurement strength.

\section{Dynamics of observables under noise}\label{app:B}
\begin{figure}[t!]
\captionsetup[subfigure]{}
\begin{center}
\includegraphics[scale = 0.9]{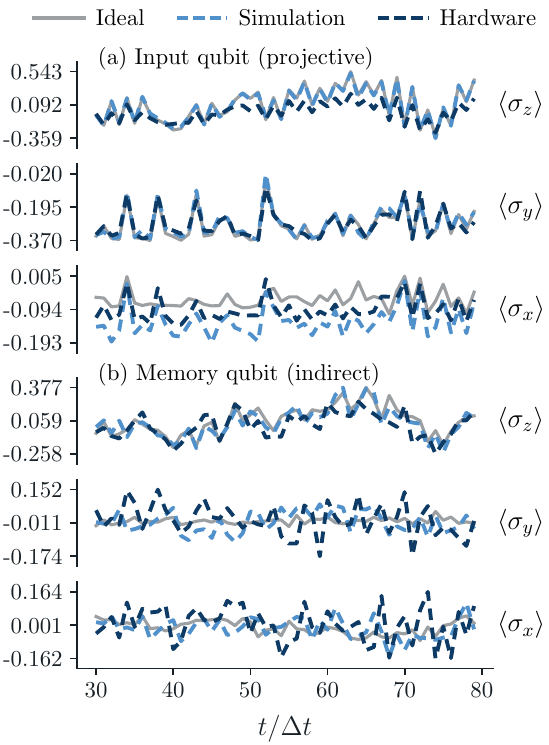}
\caption{
Comparison of expectation values obtained from the spin model, the circuit simulation and the hardware-tailored realization for (a) the input qubit under projective measurement and for (b) a memory qubit under indirect measurement. 
}\label{Fig:dyn_app}
\end{center}
\end{figure}

To complement Figs.~\ref{Fig:2} and ~\ref{Fig:3}, we present an example of the dynamics of the observables in the hardware-tailored experiment here. Figure \ref{Fig:dyn_app} compares the spin model (gray continuous line), the circuit simulation with Troterrization and shot noise (light blue dashed line) and the hardware experiment (dark blue dashed line). The time window is taken from the STM task data, and all parameters coincide with those in the main text.

As expected, the circuit simulation approximates the analog spin model more accurately, with relatively high levels of noise for the observables $\langle \hat{\sigma}^{x} \rangle$ and $\langle \hat{\sigma}^{y} \rangle$ of the indirectly measured qubits. These results suggest two things: first, that projective measurements are less affected by hardware noise; and second, that $\langle \hat{\sigma}^{z} \rangle$ observables are more resilient to noise under the studied conditions, as already anticipated in Ref. \cite{mujal2023time} (Fig.4 (a)). 

\begin{figure}[b!]
\captionsetup[subfigure]{}
\begin{center}
\includegraphics[scale = 0.9]{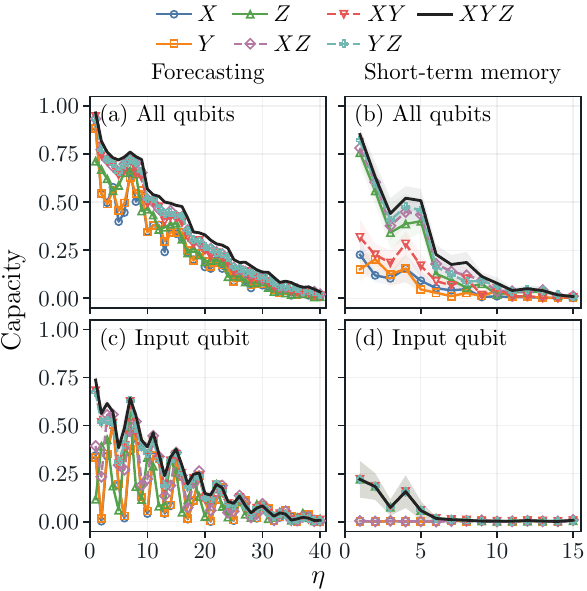}
\caption{All combinations of the output layer for Santa Fe and STM tasks.
}\label{Fig:C_combinations}
\end{center}
\end{figure}
\section{Effect of hardware noise on the tasks}\label{app:D}Figure \ref{Fig:dyn_app} shows the dynamics of the observables for the STM task. In this plot, one can see that projectively measured single-qubit observables in (a) are more resilient to hardware noise than indirectly measured qubits in (b), in particular for the $x$ and $y$ directions. This partly explains the differences in agreement between the hardware and simulations observed in \ref{Fig:3}(a) and \ref{Fig:3}(b): $g=2.05$ is very close to yielding projective measurements in the Santa Fe task, whereas $g=0.256$ is a weak measurement in the STM task. Figure \ref{Fig:C_combinations} tells the other part of the story. It shows $C(\eta)$ for the ideal spin model with an output layer made of different combinations of observables, with (c) and (d) focusing only on the observables of the projectively measured encoding qubit. For the Santa Fe task, Figs.~\ref{Fig:C_combinations} (a) and (c) show that, ideally, all observables contribute to the capacity, even the indirectly measured ones. Then, the effect of hardware noise is diluted since it is small (because of the nearly-projective measurements) and similar for all observables. However, Fig.~\ref{Fig:C_combinations} (b) shows that, ideally, the contributions to the STM capacity mainly rely on the $z$ direction, with significant contributions also coming from the $x$ and $y$ directions. The fact that most linear memory falls in the $z$ direction can be understood in terms of the expressivity analysis of the input encoding, as studied in \cite{mujal2021analytical}. Importantly, the $x$ and $y$ contributions are coming from the weakly measured qubits, as Fig.~\ref{Fig:C_combinations} (d) only shows a small capacity coming from the $z$ direction of the encoding qubit. Together, these observations explain the different sensitivity of the two tasks to hardware noise observed in Fig.~\ref{Fig:3}.

\end{document}